\documentclass [12pt]{article}
\usepackage{graphicx,amssymb,amsfonts,latexsym,amsmath,amsthm,times}
\usepackage{epsfig}
\usepackage{fancyhdr}
\usepackage{color}
\usepackage[english]{babel}
\numberwithin{equation}{section}

\newcommand{\operL}{\hat{\mathcal{L}}}
\newcommand{\calH}{\mathcal{H}}

\begin{document}

\footnotesize {\flushleft \mbox{\bf \textit{Math. Model. Nat.
Phenom.}}}
 \\
\mbox{\textit{{\bf Vol. 10, No. 1, 2015, pp. 10-20}}}

\thispagestyle{plain}

\vspace*{2cm} \normalsize \centerline{\Large \bf Noise-Produced Patterns in Images Constructed }
\vspace*{0.4cm} \centerline{\Large \bf from Magnetic Flux Leakage Data}

\vspace*{1cm}

\centerline{\bf Denis S.\ Goldobin$^{a,b,c,}$\footnote{Corresponding
author. E-mail: Denis.Goldobin@gmail.com},
                Anastasiya V.\ Pimenova$^{a}$,
                Jeremy Levesley$^b$,}
\vspace*{0.2cm}
\centerline{\bf Peter Elkington$^d$,
                Mark Bacciarelli$^d$}

\vspace*{0.5cm}

\centerline{$^a$ Institute of Continuous Media Mechanics, UB RAS, Perm 614013, Russia}

\centerline{$^b$ Department of Mathematics, University of Leicester, Leicester LE1~7RH, UK}

\centerline{$^c$ Department of Theoretical Physics, Perm State University, Perm 614990, Russia}

\centerline{$^d$ Weatherford, East Leake, Loughborough LE12~6JX, UK}


\vspace*{1cm}

\noindent {\bf Abstract.}
Magnetic flux leakage measurements help identify the position, size and shape of corrosion-related defects in steel casings used to protect boreholes drilled into oil and gas reservoirs. Images constructed from magnetic flux leakage data contain patterns related to noise inherent in the method. We investigate the patterns and their scaling properties for the case of delta-correlated input noise, and consider the implications for the method's ability to resolve defects. The analytical evaluation of the noise-produced patterns is made possible by model reduction facilitated by large-scale approximation. With appropriate modification, the approach can be employed to analyze noise-produced patterns in other situations where the data of interest are not measured directly, but are related to the measured data by a complex linear transform involving integrations with respect to spatial coordinates.

\vspace*{0.5cm}




\vspace*{1cm}


\setcounter{equation}{0}
\section{Introduction}
A common challenge in engineering is the need to infer properties of interest from indirect measurements. The direct measurement of the property of interest may be possible but expensive or inefficient compared to an indirect measurement---an example being the determination of methane hydrate saturation profiles in seafloor sediments, which can be done efficiently and cheaply by indirect means by measuring water salinity profiles from which saturations can be determined (e.g., see~\cite{Davie-Buffett-2001,Goldobin-2013}). The challenge is particularly widespread in methods employing acoustic and electromagnetic wave scattering (e.g., see~\cite{Cagniard-1953,Eberhard-1982,Hollingworth-etal-2000}), and in the deformation of static fields due to a medium's susceptibility to these fields. An example of the latter is the magnetic flux leakage (MFL) method used to assess casing integrity in boreholes drilled into rock formations for the exploration and evaluation of oil and gas resources (e.g., see~\cite{Lord-Hwang-1977,Sharar-Cuthill-Edwards-2008,Pimenova-etal-2015}). In these cases the profiles of interest are complex transforms of the measured data involving integrations with respect to spatial coordinates. These transforms can produce non-trivial and unexpected responses in the reconstructed (output) data due to noise inherent to the measurement. Even for uncorrelated input data the noise-produced patterns in reconstructed data can be correlated, smooth and indistinguishable from the actual patterns in an idealistic noise-free situation. (In some cases, the appearance of smooth, noise-produced patterns can be treated as a result of noise accumulation in the reconstructed data.) Therefore noise-produced patterns cannot simply be filtered out by means of some data processing procedure (which would be the case for uncorrelated noise). Such noise-induced patterns can significantly influence the utility of the reconstruction procedure, and knowledge of the properties of these patterns is therefore important, and serves as an valuable guide in the development of data analysis algorithms and in future device design.

\begin{figure}[!t]
\center{
\includegraphics[width=0.43\textwidth]%
 {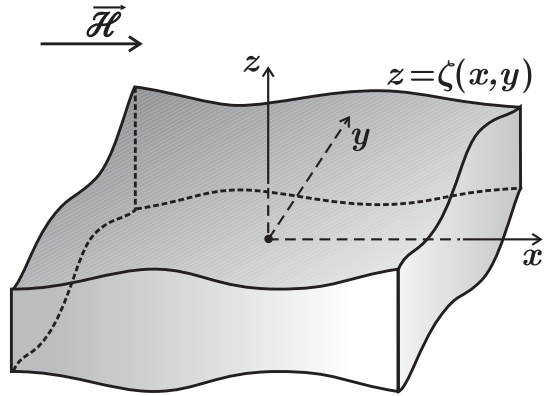}
}

  \caption{
Ferromagnetic layer of non-uniform thickness in the field $\vec{\calH}$ and the coordinate frame}
  \label{fig1}
\end{figure}

In this paper we suggest and implement an approach to comprehensively analyse noise-produced patterns in MFL-reconstructed data for different device designs. The problem set-up corresponds to measurements made with modern devices designed for the MFL inspection of wellbore casings described for example in~\cite{Sharar-Cuthill-Edwards-2008}. Our approach is based on the mathematical methodology developed in~\cite{Pimenova-etal-2015} within the framework of large-scale approximation, which is relevant for the case of corrosive damage. Such model reduction provides the opportunity for a comprehensive analysis whose results remain qualitatively valid for the general case beyond large-scale approximation. In the analysis of inaccuracies, it is generally sufficient to have estimates of errors. A similar approach can be employed for various data reconstruction tasks.

The paper is organized as follows. In Section~2 we introduce the basic mathematical model we use. In Section~3 we evaluate the patterns in the reconstructed data produced by $\delta$-correlated noise in the measured magnetic field data. In Section~4 we analyse the scaling properties of the noise-produced patterns for various device designs, compare the cases of different designs, and discuss the role of the defect geometry for their resolvability against the background of noise-produced patterns. Conclusions are drawn in Section~5.

\setcounter{equation}{0}
\section{Magnetic field in the presence of large-scale defects}
\label{sec:mathmod}
The system under consideration is shown in Figure~\ref{fig1}; it comprises a ferromagnetic layer subject to an external magnetic field parallel to the layer middle-plane; the $x$-axis is directed along $\vec\calH$, and the $z$-axis is perpendicular to the layer middle-plane. Within the framework of large-scale approximation, the magnetic field within the ferromagnetic layer is nearly homogeneous along the $z$-axis and is parallel to the layer-middle plane, and thus can be represented by a gradient of potential $\phi(x,y)$, which is a function of two coordinates $(x,y)$ only: $\vec{H}^\mathrm{(in)}=-\nabla\phi(x,y)$. To the leading order, this potential obeys the equation~\cite{Pimenova-etal-2015}
\begin{equation}
\label{eq101}
\nabla_2\cdot\left(e^{-\sigma(x,y)}\nabla_2\phi(x,y)\right)=0\,,
\end{equation}
where subscript `2' indicates that $\nabla_2$ operates in the two-dimensional space of $(x,y)$. The function $\sigma(x,y)$ describes the relative loss of the ferromagnetic material; the layer thickness profile
$$
2\zeta(x,y)=2\zeta_0e^{-\sigma(x,y)}\,,
$$
where $2\zeta_0$ is the thickness of the undamaged layer.

Outside of the layer the magnetic field is the 3-d gradient of a harmonic potential;
\begin{equation}
\label{eq102}
\vec{H}^\mathrm{(ext)}=-\nabla\varphi(x,y,z)\,,
\end{equation}
\begin{equation}
\label{eq103}
\Delta\varphi(x,y,z)=0\,,
\end{equation}
with the boundary condition on the ferromagnet surface
$$
\varphi(x,y,z=\zeta(x,y))=\phi(x,y)
$$
and $(-\nabla\varphi)_{z\to\pm\infty}=\vec{\calH}$\,. Within the framework of large-scale approximation, to the leading order, one can translate the boundary condition for $\varphi$ from the layer surface to $z=0$:
\begin{equation}
\label{eq104}
\varphi(x,y,0)=\phi(x,y)\,.
\end{equation}

By virtue of Eq.\,(\ref{eq103}), the potential $\varphi$ of the external field can be represented in the basis of exponential functions
\begin{equation}
\label{eq105}
\varphi(x,y,z)=-\calH x
+\iint\mathrm{d}k_x\,\mathrm{d}k_y\,A(k_x,k_y)\,e^{-k|z|}\,e^{i(k_xx+k_yy)}\,,
\quad k=\sqrt{k_x^2+k_y^2}\,.
\end{equation}
Here $A(k_x,k_y)e^{-k|z_\mathrm{m}|}$ is the Fourier transform of the deformation of the magnetic potential field, $\varphi+\calH x$, measured on the plane elevated by $(z_\mathrm{m}-\zeta_0)$ above the layer surface. While the field $\varphi(x,y,z_\mathrm{m})$ can be found from measurements of $H_x^\mathrm{(ext)}(x,y,z_\mathrm{m})$ or/and $H_y^\mathrm{(ext)}(x,y,z_\mathrm{m})$ by plain integration with respect to $x$ or $y$, evaluation of $\varphi(x,y,z_\mathrm{m})$ from measurements of $H_z^\mathrm{(ext)}(x,y,z_\mathrm{m})$ is non-trivial; the relationship between $\varphi(x,y,z_\mathrm{m})$ and $H_z^\mathrm{(ext)}(x,y,z_\mathrm{m})$ is non-local. For this relationship, one can rely on Eq.\,(\ref{eq105}) and its $z$-derivative;
\begin{equation}
\label{eq106}
H_z^\mathrm{(ext)}(x,y,z)=\iint\mathrm{d}k_x\,\mathrm{d}k_y\,k\,A(k_x,k_y)\,e^{-k|z|}\,e^{i(k_xx+k_yy)}\,,
\quad k=\sqrt{k_x^2+k_y^2}\,.
\end{equation}
Thus, one can introduce a non-local operator $\operL$ which transforms $\varphi(x,y,z_\mathrm{m})$ (a function of two variables $(x,y)$) into $H_z^\mathrm{(ext)}(x,y,z_\mathrm{m})$ (another function of two variables $(x,y)$):
\begin{equation}
\label{eq107}
-\frac{\partial}{\partial z}\varphi(x,y,z_\mathrm{m})=\operL\varphi(x,y,z_\mathrm{m})\,.
\end{equation}
This transform is correctly defined for any harmonic function vanishing for $z\to\pm\infty$, and in the Fourier space it is a multiplication by $k=|\vec{k}|$. For trivial conditions on the boundaries limiting the system in the $x$- and $y$-directions, the inverse operator $\operL^{-1}$ is well defined; the operator $\operL$ is a homeomorphism.

The aim of our work is to establish smooth large-scale patterns in reconstructed data produced by the noise in measurements. Hence, the approximate calculation of these patterns will suffice, and we may make a further model reduction for Eq.\,(\ref{eq101}) and the system under consideration. Specifically, we make this reduction for $z_\mathrm{m}$ small compared to the scale of defects, $\phi(x,y)\approx\varphi(x,y,z_\mathrm{m})$. Hence, Eq.\,(\ref{eq101}) yields an equation for calculation of $\sigma(x,y)$ from measurable derivatives of $\phi(x,y)$: $\nabla_2\phi\cdot\nabla_2\sigma=\Delta_2\phi$. To the leading order, $\vec{H}^\mathrm{(ext)}(z_\mathrm{m})\approx\vec{\calH}$; therefore, Eq.\,(\ref{eq101}) can be approximately rewritten as
\begin{equation}
\label{eq108}
\calH\frac{\partial\sigma}{\partial x}=-\Delta_2\varphi(x,y,z_\mathrm{m})
 =-\left.\frac{\partial H_z^\mathrm{(ext)}}{\partial z}\right|_{z=z_\mathrm{m}}.
\end{equation}

\setcounter{equation}{0}
\section{Delta-correlated noise in measurements}
\label{sec:deltacorr}
Let us consider noise in the measurement data;
\begin{equation}
\label{eq201}
f(x,y)=f_0(x,y)+a\,\xi(x,y)\,,
\end{equation}
where $f(x,y)$ is the measured field, $f_0(x,y)$ is the actual field without distortion by noise in measurements, and $a\,\xi(x,y)$ is the measurement noise with amplitude $a$ and $\xi(x,y)$ normalized as specified below in the text. We assume that the noise at the neighbouring measurement points to be uncorrelated (bias in measurements can be removed by calibration of the measuring sensor), that is $\xi(x,y)$ can be assumed to be $\delta$-correlated. We choose the following normalization: $\langle\xi(x,y)\,\xi(x',y')\rangle=4\delta(x-x')\,\delta(y-y')$. Here and hereafter, $\langle\dots\rangle$ means averaging over the noise realizations. The amplitude $a$ of normalized noise is determined by the variance of the measurement error $\varepsilon_f^\mathrm{(meas)}=f^\mathrm{(meas)}-f_0$;
\begin{equation}
\label{eq202}
\langle(\varepsilon_f^\mathrm{(meas)})^2\rangle=a^2\langle\xi^2\rangle=a^2\frac{4}{h_x\,h_y}\,,
\end{equation}
where $h_x$ and $h_y$ are the distances between the measurement points in the $x$- and $y$-directions, respectively.

\subsection{Measuring $f(x,y)=\partial H_z/\partial x$}
Let us first consider the case of measuring $f(x,y)=\partial H_z^\mathrm{(ext)}/\partial x$, which is routinely performed by devices in common usage~\cite{Sharar-Cuthill-Edwards-2008}. According to Eq.\,(\ref{eq108}), the reconstructed profile
\begin{align}
\sigma(x,y)&=\frac{1}{\calH}\int^x\mathrm{d}x'\,\operL\int^{x'}\mathrm{d}x'' f(x'',y)
\nonumber\\[5pt]
&=\frac{1}{\calH}\operL\int^x\mathrm{d}x'\int^{x'}\mathrm{d}x''\,f(x'',y)\,.
\label{eq203}
\end{align}
Here one can change the order of operation of the integration with respect to $x$ and the operator $\operL$, which is differentiation with respect to $z$. Hence, the purely noise-produced pattern in the reconstructed profile $\sigma(x,y)$ is
\begin{equation}
\varepsilon_\sigma(x,y) =\frac{1}{\calH}\operL\int^x\mathrm{d}x'\int^{x'}\mathrm{d}x''\,\varepsilon_f(x'',y)\,.
\label{eq204}
\end{equation}

\begin{figure}[!t]

\centerline{
\begin{tabular}{ccc}
\multicolumn{3}{c}{
(a)\qquad
\includegraphics[width=0.333\textwidth]%
 {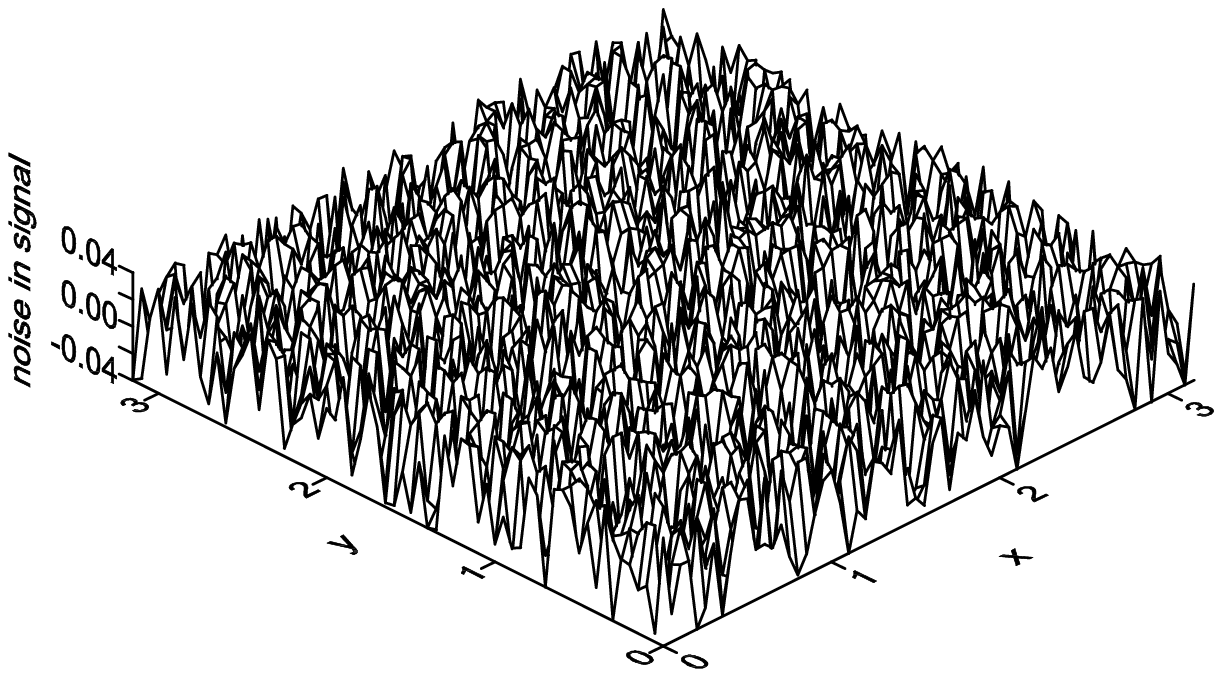}}
\\[8pt]
(b)
&
\includegraphics[width=0.333\textwidth]%
 {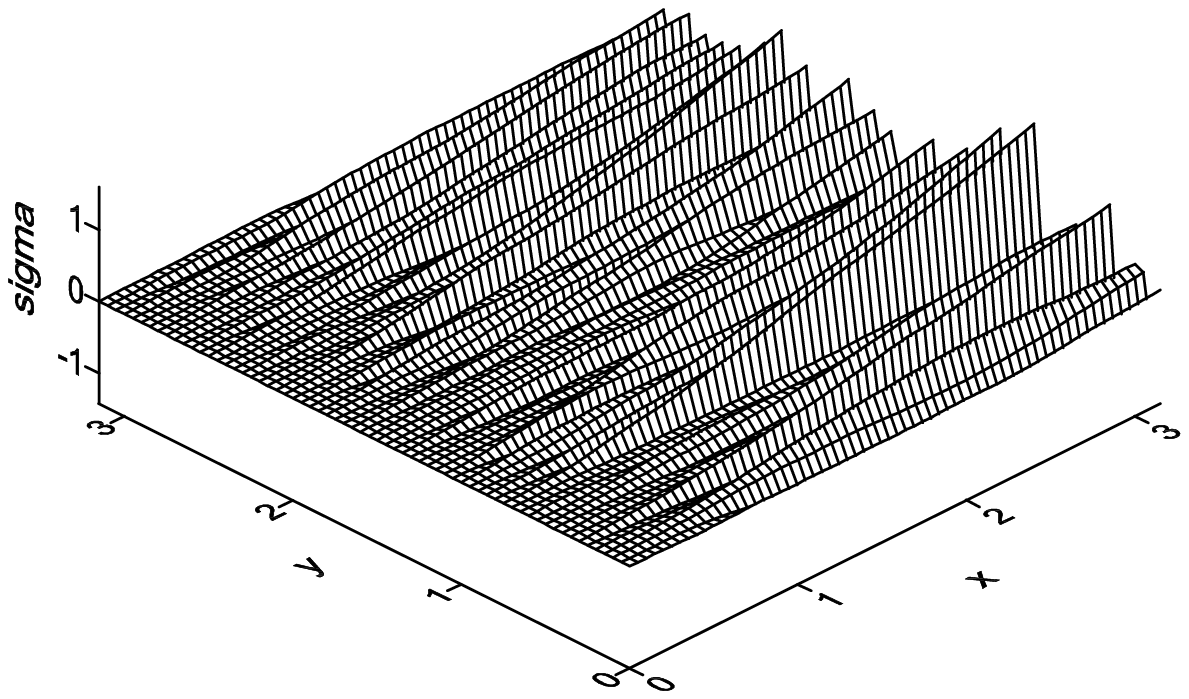}\qquad
&\qquad
\includegraphics[width=0.333\textwidth]%
 {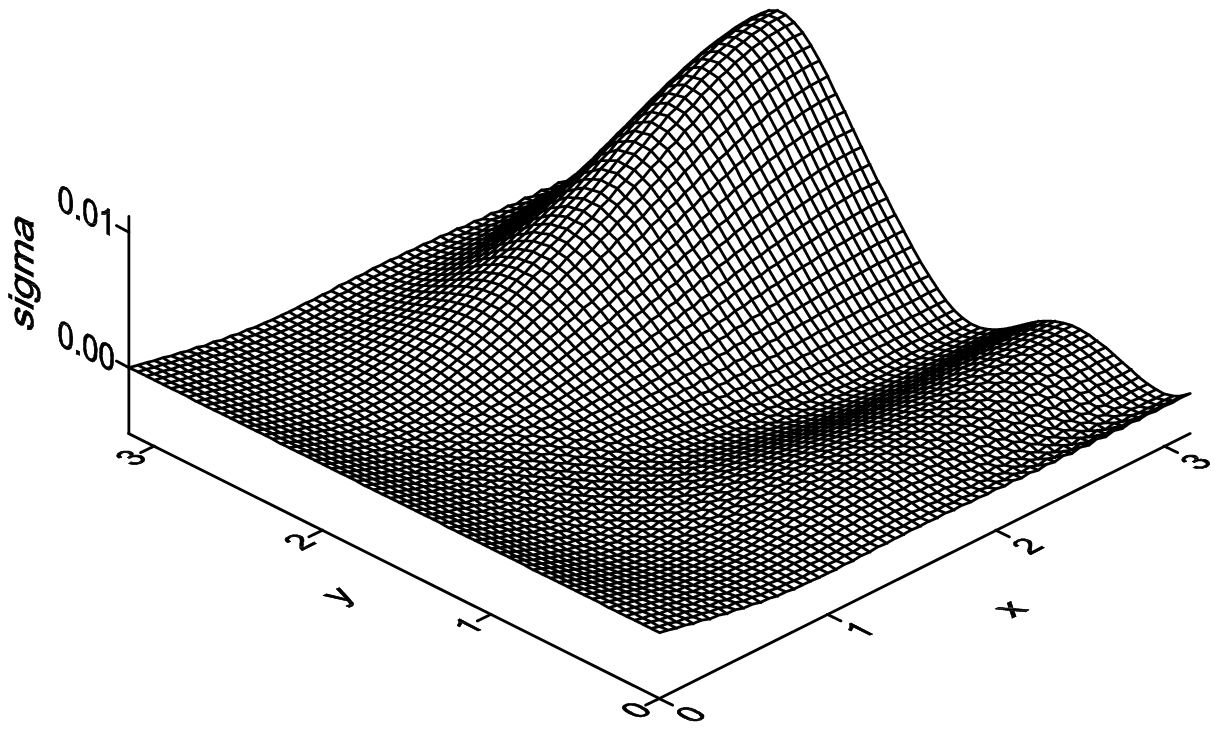}
\\[8pt]
(c)
&
\includegraphics[width=0.333\textwidth]%
 {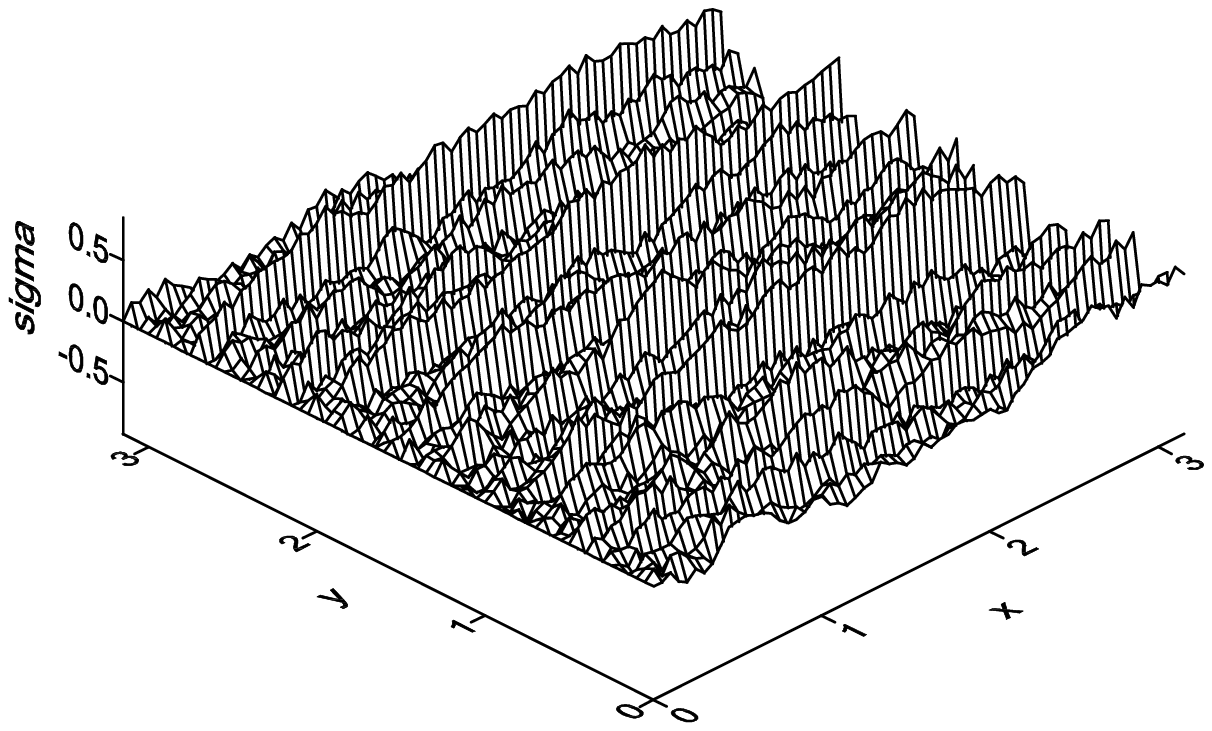}\qquad
&\qquad
\includegraphics[width=0.333\textwidth]%
 {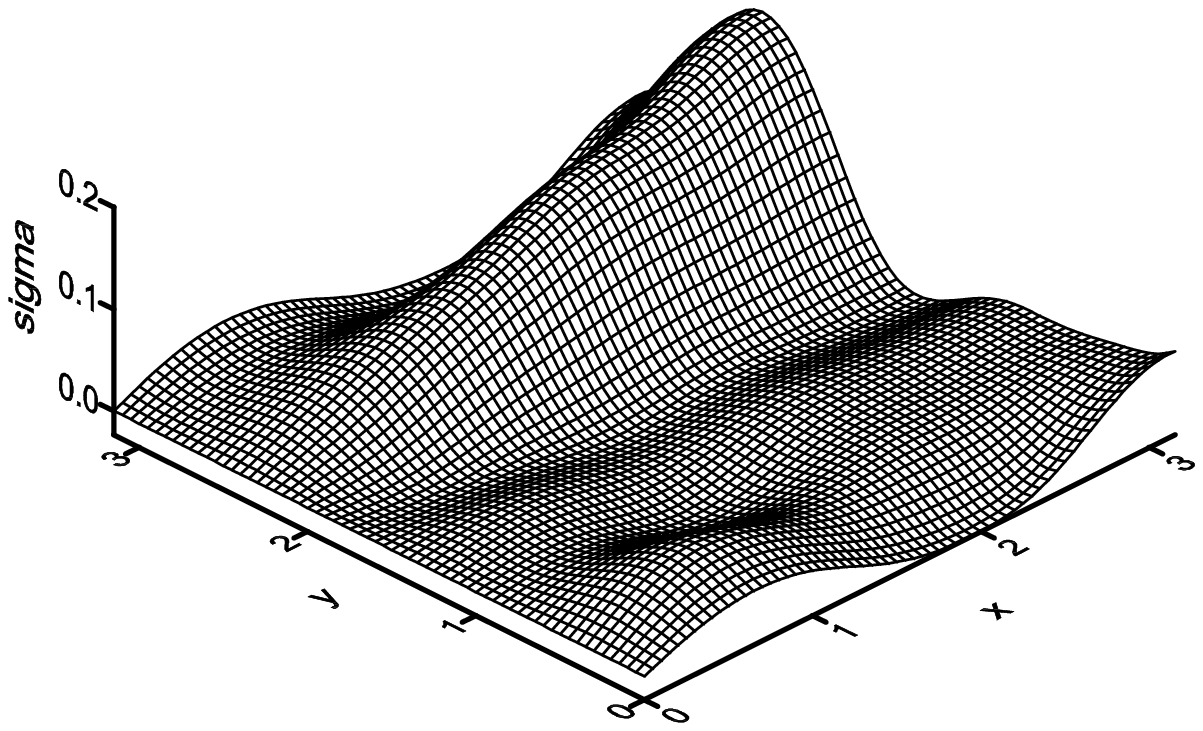}
\\[8pt]
(d)
&
\includegraphics[width=0.333\textwidth]%
 {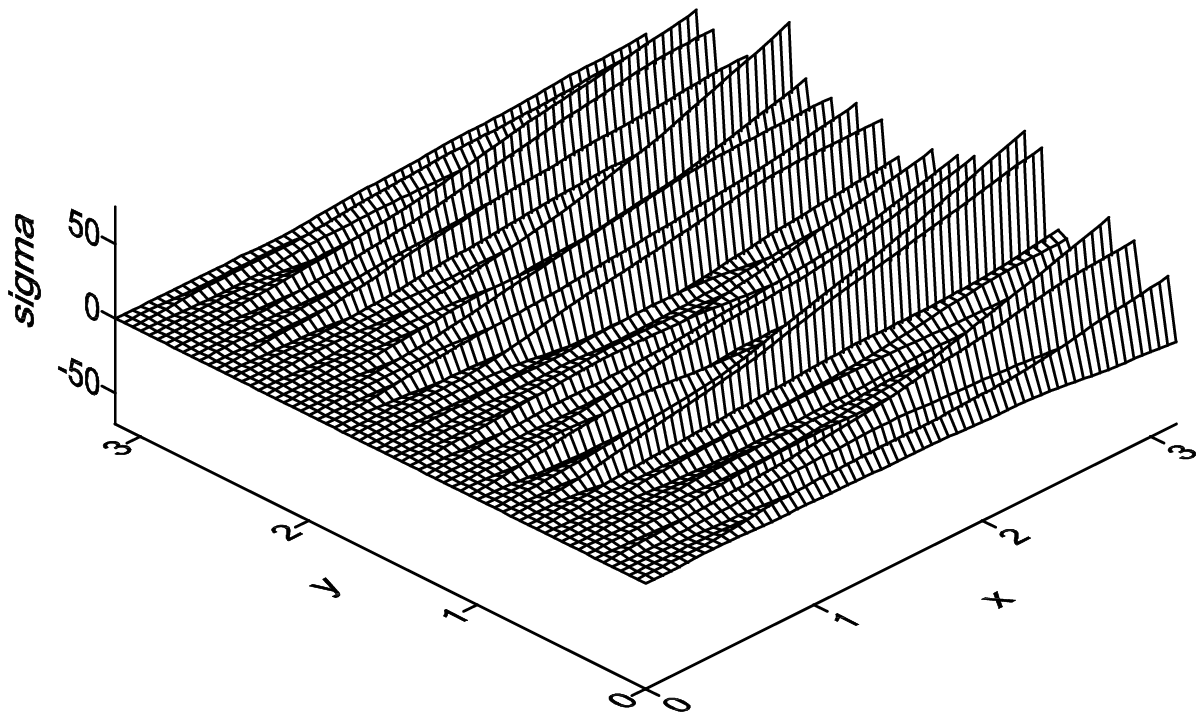}\qquad
&\qquad
\includegraphics[width=0.333\textwidth]%
 {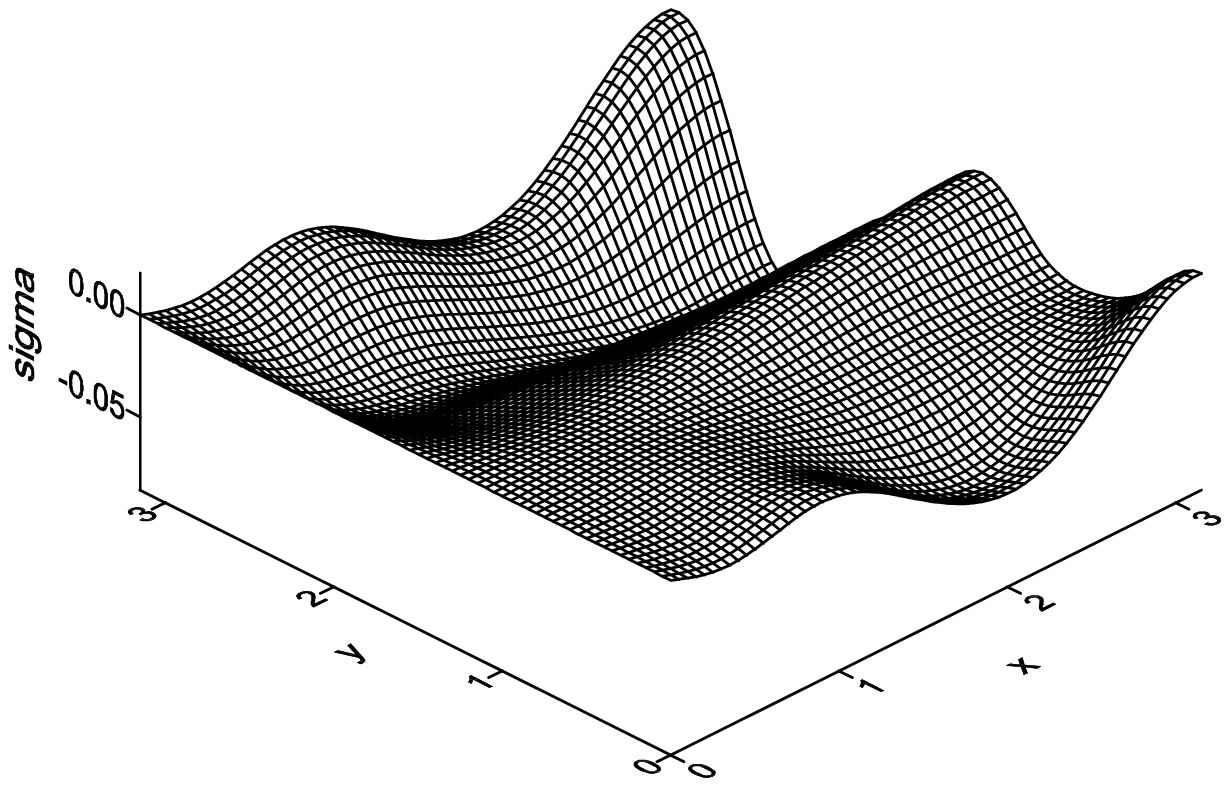}
\\[8pt]
(e)
&
\includegraphics[width=0.333\textwidth]%
 {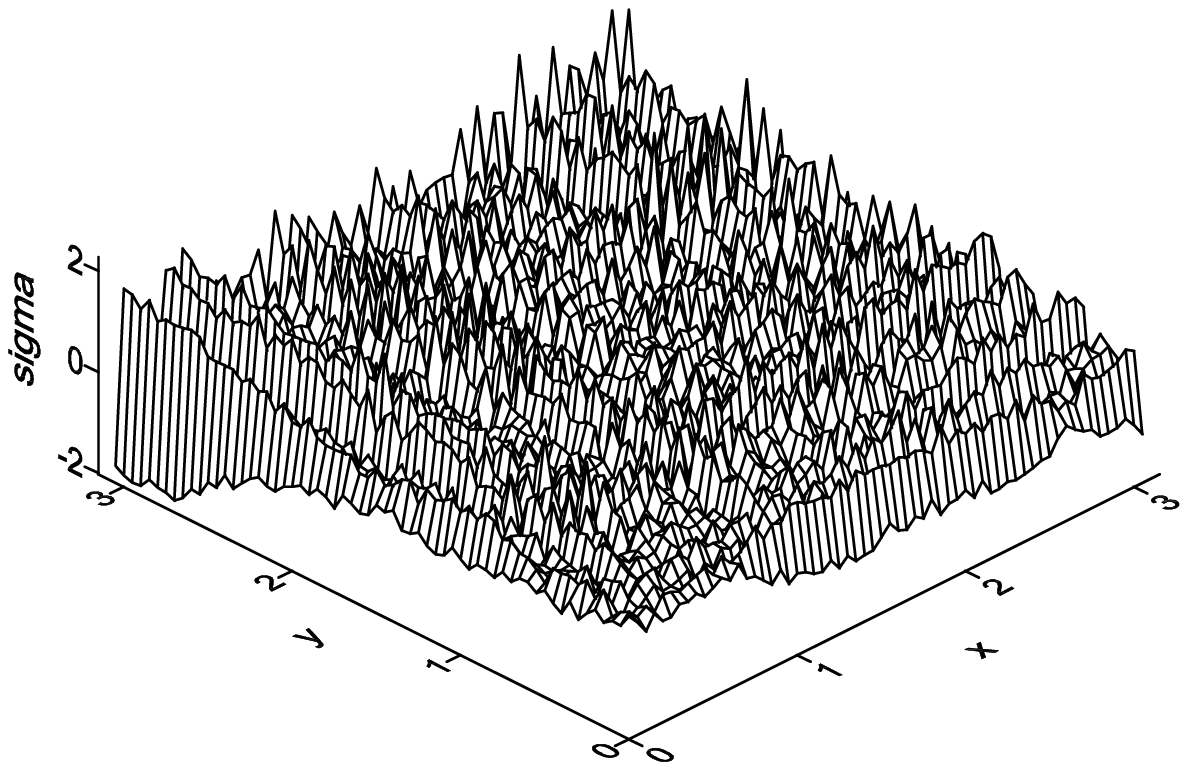}\qquad
&\qquad
\includegraphics[width=0.333\textwidth]%
 {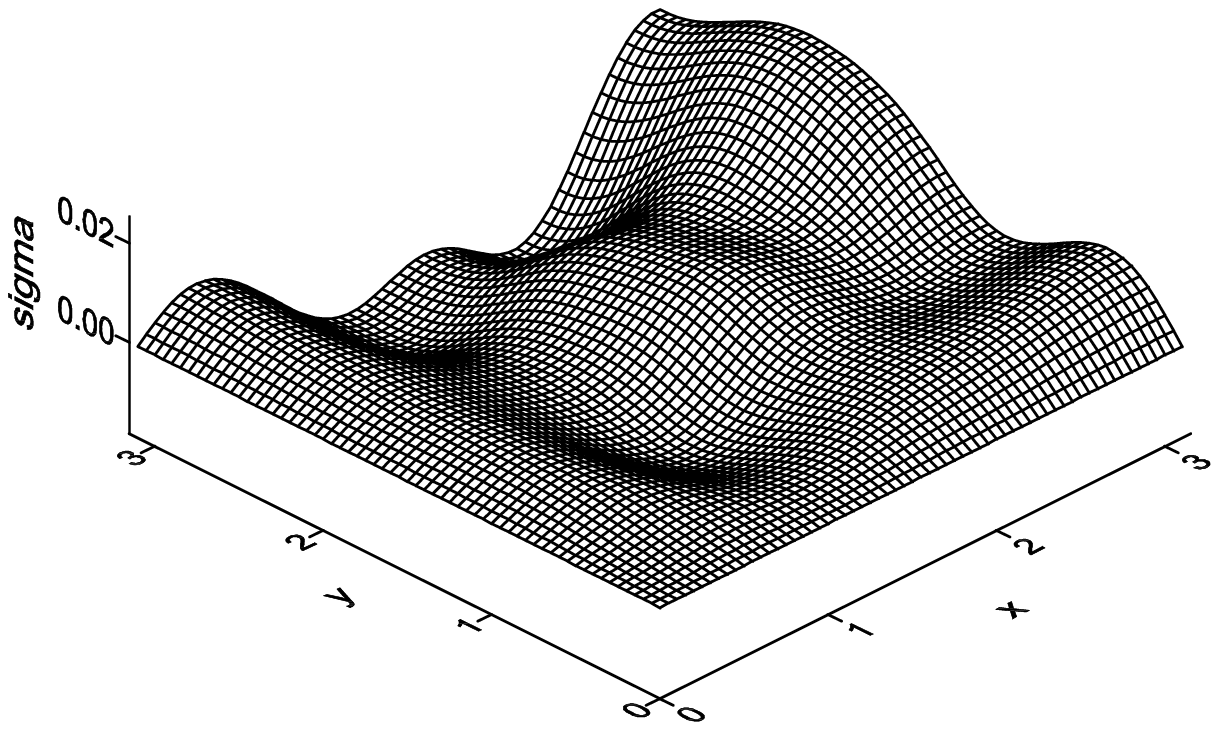}
\end{tabular}
}

  \caption{
Sample realization of noise $(a/\calH)\xi(x,y)$~(a), and corresponding reconstructed profiles $\sigma$ (left column), and profiles $\sigma$  with small-scale oscillations filtered-out (right column). (b):~$f(x,y)=(\partial H_z/\partial x)$\,, (c):~$f(x,y)=H_z$\,, (d):~$f(x,y)=H_x$\,, (e):~$f(x,y)=H_y$\,. The smoothed profiles are a consequence of noise and should not be misinterpreted as real defects.}
  \label{fig2}
\end{figure}

In contrast to the noise in $f(x,y)$, which is $\delta$-correlated, the noise-produced pattern $\varepsilon_\sigma(x,y)$ is a result of two integrations and the action of non-local operator $\operL$\,; therefore, this pattern can have a smooth component with a finite correlation length which will not be easily recognisable against the background of the actual noise-free profile $\sigma_0(x,y)$. In Figure~\ref{fig2}b one can see the profile $\varepsilon_\sigma(x,y)$ and its smooth component calculated for $f(x,y)=a\xi(x,y)$ plotted in Figure~\ref{fig2}a. We need to evaluate the characteristic magnitude of these smoothed profiles on the scale $L_x\times L_y$\,.

To calculate the characteristic magnitude of $\varepsilon_\sigma$, one can consider the average value of $\varepsilon_\sigma$ over the area $S=L_x\times L_y$;
\begin{equation}
\label{eq205}
\overline{\varepsilon_{\sigma}(x,y)}= \frac{1}{L_xL_y}\iint_{S}\varepsilon_\sigma(x,y)\,\mathrm{d}x\,\mathrm{d}y\,.
\end{equation}
As $\xi(x,y)$ is $\delta$-correlated, its integral is a Gaussian random variable by virtue of the central limit theorem. It is more convenient to evaluate the magnitude of $\overline{\operL^{-1}\varepsilon_\sigma(x,y)}$ and then assess how $\operL$ influences the magnitude of its $(L_x\times L_y)$-scale component. According to the central limit theorem, the area-average
\begin{equation}
\label{eq206}
\overline{\operL^{-1}\varepsilon_{\sigma}(x,y)}= \frac{1}{L_xL_y}\iint_{S}\operL^{-1}\varepsilon_\sigma(x,y)\,\mathrm{d}x\,\mathrm{d}y
\end{equation}
is a Gaussian random variable with variance
\begin{align}
\left\langle\left(\overline{\operL^{-1}\varepsilon_{\sigma}}\right)^2\right\rangle&= \frac{1}{L_x^2\,L_y^2}\left\langle\left(\iint_{S}\operL^{-1}\varepsilon_\sigma(x,y)\,\mathrm{d}x\,\mathrm{d}y\right)^2\right\rangle
\nonumber\\[5pt]
&=\frac{1}{L_x^2\,L_y^2\,\calH^2}\left\langle\int_0^{L_x}\!\!\mathrm{d}x_1\int_0^{L_y}\!\!\mathrm{d}y_1
\int_0^{x_1}\!\!\mathrm{d}x'_1\int_0^{x'_1}\!\!\mathrm{d}x''_1\,\varepsilon_f(x''_1,y_1)
\right.
\nonumber\\
&\qquad\qquad\qquad
\left.\times\int_0^{L_x}\!\!\mathrm{d}x_2\int_0^{L_y}\!\!\mathrm{d}y_2
\int_0^{x_2}\!\!\mathrm{d}x'_2\int_0^{x'_2}\!\!\mathrm{d}x''_2\,\varepsilon_f(x''_2,y_2)
\right\rangle
\nonumber\\[5pt]
&=\frac{1}{L_x^2\,L_y^2\,\calH^2}\int_0^{L_x}\!\!\mathrm{d}x_1\int_0^{L_y}\!\!\mathrm{d}y_1
\int_0^{x_1}\!\!\mathrm{d}x'_1\int_0^{x'_1}\!\!\mathrm{d}x''_1\,
\nonumber\\
&\qquad\qquad\quad
\times\int_0^{L_x}\!\!\mathrm{d}x_2\int_0^{L_y}\!\!\mathrm{d}y_2
\int_0^{x_2}\!\!\mathrm{d}x'_2\int_0^{x'_2}\!\!\mathrm{d}x''_2
\left\langle\varepsilon_f(x''_1,y_1)\,\varepsilon_f(x''_2,y_2)\right\rangle
\nonumber\\[5pt]
&=\frac{4a^2}{L_x^2\,L_y^2\,\calH^2}\int_0^{L_x}\!\!\mathrm{d}x_1\int_0^{L_y}\!\!\mathrm{d}y_1
\int_0^{x_1}\!\!\mathrm{d}x'_1\int_0^{x'_1}\!\!\mathrm{d}x''_1\,
\nonumber\\
&\qquad\qquad\quad
\times\int_0^{L_x}\!\!\mathrm{d}x_2\int_0^{L_y}\!\!\mathrm{d}y_2
\int_0^{x_2}\!\!\mathrm{d}x'_2\int_0^{x'_2}\!\!\mathrm{d}x''_2\,
\delta(x''_1-x''_2)\,\delta(y_1-y_2)
\nonumber\\[5pt]
&=\frac{4a^2}{L_x^2\,L_y^2\,\calH^2}\int_0^{L_x}\!\!\mathrm{d}x_1
\int_0^{x_1}\!\!\mathrm{d}x'_1\int_0^{x'_1}\!\!\mathrm{d}x''_1\,
\int_0^{L_x}\!\!\mathrm{d}x_2
\int_0^{x_2}\!\!\mathrm{d}x'_2\int_0^{x'_2}\!\!\mathrm{d}x''_2\,
L_y\,\delta(x''_1-x''_2)
\nonumber\\[5pt]
&=\frac{4a^2}{L_x^2\,L_y^2\,\calH^2}\frac{L_y\,L_x^5}{20}\,.
\label{eq207}
\end{align}

According to Eq.\,(\ref{eq106}), the action of the operator $\operL$ in Fourier space is multiplication by the wavenumber $k=(k_x^2+k_y^2)^{1/2}$. Hence, for the mode with scale $L_x\times L_y$---i.e., $k_x=\pi/L_x$ and $k_y=\pi/L_y$---action of $\operL$ is equivalent to multiplication by the factor $\pi(L_x^{-2}+L_y^{-2})^{1/2}$. Let $a_\sigma$ be the characteristic magnitude of the noise-produced profile on the scale $L_x\times L_y$. Then Eq.\,(\ref{eq207}) yields the representation
\begin{align}
a_\sigma^{(\partial{H_z}/\partial{x})}=\sqrt{\left\langle\left(\overline{\varepsilon_{\sigma}}\right)^2\right\rangle}
&\approx\frac{\pi}{\sqrt{5}}\frac{a}{\calH}\frac{1}{\sqrt{S}}\frac{L_x}{L_y}\sqrt{L_x^2+L_y^2}
\nonumber\\[5pt]
&=\frac{\pi\sqrt{\langle(\varepsilon_f^\mathrm{(meas)})^2\rangle\,h_x\,h_y}}{2\sqrt{5}\,\calH}\frac{1}{\sqrt{S}}\frac{L_x}{L_y}\sqrt{L_x^2+L_y^2}\,.
\label{eq208}
\end{align}
Here we explicitly distinguish the factor $1/\sqrt{S}$ which features the ``reference'' convergence law for the average value of a noisy variable with respect to the length scale $L_x$, $L_y$. One can observe not simply a slower convergence of the amplitude of the noise-produced pattern but its independence on the scale length for $L_x\sim L_y$. The error is the same on all spatial scales, suggesting that measurement $(\partial{H_z}/\partial{x})$ is equally applicable for detection of defects on different spatial scales.

\subsection{Measuring $f(x,y)=H_z$, $H_x$ or $H_y$}
Alternatively to the previous case, some devices measure $H_z=-\partial\varphi/\partial z$, $H_x=-\partial\varphi/\partial x$ and/or $H_y=-\partial\varphi/\partial y$. In Figure~\ref{fig2} one can see sample profiles $\varepsilon_\sigma(x,y)$ for the cases of measuring different components of $\vec{H}$. Similarly to Eq.\,(\ref{eq204}) for the case of $f=\partial H_z/\partial x$, Eq.\,(\ref{eq108}) yields
\begin{align}
\varepsilon_\sigma^{(H_z)}(x,y)
&=-\frac{1}{\calH}\operL\int^x\mathrm{d}x'\,\varepsilon_f^{(H_z)}(x',y)\,,
\label{eq209}
\\[5pt]
\varepsilon_\sigma^{(H_x)}(x,y) &=-\frac{1}{\calH}\operL^2\int^x\mathrm{d}x'\int^{x'}\mathrm{d}x''\,\varepsilon_f^{(H_x)}(x'',y)\,,
\label{eq210}
\\[5pt]
\varepsilon_\sigma^{(H_y)}(x,y)
&=-\frac{1}{\calH}\operL^2\int^x\mathrm{d}x'\int^{y}\mathrm{d}y'\,\varepsilon_f^{(H_y)}(x',y')\,.
\label{eq211}
\end{align}
Area-averages of $\varepsilon_{\sigma}^{(H_z)}$, $\varepsilon_{\sigma}^{(H_x)}$ and $\varepsilon_{\sigma}^{(H_y)}$ are Gaussian random variables, the variances of which are related to the following variances:
\begin{align}
&\left\langle\left(\overline{\operL^{-1}\varepsilon_{\sigma}^{(H_z)}}\right)^2\right\rangle =\frac{1}{L_x^2\,L_y^2\,\calH^2}\int_0^{L_x}\!\!\mathrm{d}x_1\int_0^{L_y}\!\!\mathrm{d}y_1
\int_0^{x_1}\!\!\mathrm{d}x'_1
\nonumber\\
&\quad
\times\int_0^{L_x}\!\!\mathrm{d}x_2\int_0^{L_y}\!\!\mathrm{d}y_2
\int_0^{x_2}\!\!\mathrm{d}x'_2
\left\langle\varepsilon_f^{(H_z)}(x'_1,y_1)\,\varepsilon_f^{(H_z)}(x'_2,y_2)\right\rangle
=\frac{4a^2}{L_x^2\,L_y^2\,\calH^2}\frac{L_y\,L_x^3}{3}\,,
\label{eq212}
\\[10pt]
&\left\langle\left(\overline{\operL^{-2}\varepsilon_{\sigma}^{(H_x)}}\right)^2\right\rangle =\frac{1}{L_x^2\,L_y^2\,\calH^2}\int_0^{L_x}\!\!\mathrm{d}x_1\int_0^{L_y}\!\!\mathrm{d}y_1
\int_0^{x_1}\!\!\mathrm{d}x'_1\int_0^{x'_1}\!\!\mathrm{d}x''_1
\nonumber\\
&\quad
\times\int_0^{L_x}\!\!\mathrm{d}x_2\int_0^{L_y}\!\!\mathrm{d}y_2
\int_0^{x_2}\!\!\mathrm{d}x'_2\int_0^{x'_2}\!\!\mathrm{d}x''_2
\left\langle\varepsilon_f^{(H_x)}(x''_1,y_1)\,\varepsilon_f^{(H_x)}(x''_2,y_2)\right\rangle
=\frac{4a^2}{L_x^2\,L_y^2\,\calH^2}\frac{L_y\,L_x^5}{20}\,,
\label{eq213}
\\[10pt]
&\left\langle\left(\overline{\operL^{-2}\varepsilon_{\sigma}^{(H_y)}}\right)^2\right\rangle =\frac{1}{L_x^2\,L_y^2\,\calH^2}\int_0^{L_x}\!\!\mathrm{d}x_1\int_0^{L_y}\!\!\mathrm{d}y_1
\int_0^{x_1}\!\!\mathrm{d}x'_1\int_0^{y_1}\!\!\mathrm{d}y'_1
\nonumber\\
&\quad
\times\int_0^{L_x}\!\!\mathrm{d}x_2\int_0^{L_y}\!\!\mathrm{d}y_2
\int_0^{x_2}\!\!\mathrm{d}x'_2\int_0^{y_2}\!\!\mathrm{d}y'_2
\left\langle\varepsilon_f^{(H_y)}(x'_1,y'_1)\,\varepsilon_f^{(H_y)}(x'_2,y'_2)\right\rangle
=\frac{4a^2}{L_x^2\,L_y^2\,\calH^2}\frac{L_y^3\,L_x^3}{9}\,.
\label{eq214}
\end{align}

Finally, the characteristic magnitudes of patterns on the scale $L_x\times L_y$ are
\begin{align}
a_\sigma^{(H_z)}&\approx\frac{\pi\sqrt{\langle(\varepsilon_f^\mathrm{(meas)})^2\rangle\,h_x\,h_y}}{\sqrt{3}\,\calH}
 \frac{1}{\sqrt{S}}\frac{L_x}{L_y}\frac{\sqrt{L_x^2+L_y^2}}{L_x}\,,
\label{eq215}
\\[5pt]
a_\sigma^{(H_x)}&\approx\frac{\pi\sqrt{\langle(\varepsilon_f^\mathrm{(meas)})^2\rangle\,h_x\,h_y}}{2\sqrt{5}\,\calH}
 \frac{1}{\sqrt{S}}\frac{L_x}{L_y}\frac{L_x^2+L_y^2}{L_y^2}\,,
\label{eq216}
\\[5pt]
a_\sigma^{(H_y)}&\approx\frac{\pi\sqrt{\langle(\varepsilon_f^\mathrm{(meas)})^2\rangle\,h_x\,h_y}}{3\,\calH}
 \frac{1}{\sqrt{S}}\left(\frac{L_x}{L_y}+\frac{L_y}{L_x}\right)\,.
\label{eq217}
\end{align}
It is noteworthy that, in contrast to the case of $f={\partial H_z/\partial x}$, Eqs.\,(\ref{eq215})--(\ref{eq217}) demonstrate the decay of the noise-produced patterns on large scale according to the ``reference'' law $\propto1/\sqrt{S}$.

\begin{table}
\caption{Dependence of the form-factor on geometry of defects}
\vspace{10pt}
\centerline{
\begin{tabular}{|c|c|c|c|c|}
\hline
& $\displaystyle \gamma_{(\partial{H_z}/\partial{x})}$ & $\displaystyle \gamma_{(H_z)}$ &
 $\displaystyle \gamma_{(H_x)}$ & $\displaystyle \gamma_{(H_y)}$
\\[3pt]
\hline
&&&&
\\[-8pt]
$L_x\sim L_y$ & $\displaystyle \frac{\pi}{\sqrt{10}}$ & $\displaystyle \pi\sqrt{\frac{2}{3}}$ & $\displaystyle \frac{\pi}{\sqrt{5}}$ & $\displaystyle \frac{2\pi}{3}$
\\[16pt]
\hline
&&&&
\\[-8pt]
$L_x\ll L_y$ & $\displaystyle \frac{\pi}{2\sqrt{5}}\left(\frac{L_x}{L_y}\right)^{1/2}$ & $\displaystyle \frac{\pi}{\sqrt{3}}$ & $\displaystyle \frac{\pi}{2\sqrt{5}}\frac{L_x}{L_y}$ & $\displaystyle \frac{\pi}{3}\frac{L_y}{L_x}$
\\[16pt]
\hline
&&&&
\\[-8pt]
\ $L_x\gg L_y$ \ &
\ $\displaystyle \frac{\pi}{2\sqrt{5}}\left(\frac{L_x}{L_y}\right)^{3/2}$ \ &
\quad~$\displaystyle \frac{\pi}{\sqrt{3}}\frac{L_x}{L_y}$\quad~&
\quad~$\displaystyle \frac{\pi}{2\sqrt{5}}\frac{L_x^3}{L_y^3}$\quad~&
\quad~$\displaystyle \frac{2\pi}{3}\frac{L_x}{L_y}$\quad~
\\[16pt]
\hline
\end{tabular}}
\label{tab}
\end{table}

\setcounter{equation}{0}
\section{Discussion: Scaling of noise-produced patterns, form-factor, and optimization of measurement techniques}
\label{sec:discussion}
In this section we compare 4 possible cases of $f=(\partial H_z/\partial x)$, $H_z$, $H_x$ and $H_y$, and discuss the role of the shape of defects for their resolution.

Let us consider the discrepancy between the scaling laws of amplitude of the noise-produced patterns $a_\sigma^{(\partial H_z/\partial x)}$ and $a_\sigma^{(H_j)}$: for $L_x\sim L_y\sim L$, amplitude $a_\sigma^{(\partial H_z/\partial x)}\propto L^0$ is independent of $L$, while $a_\sigma^{(H_j)}\propto L^{-1}$. For the former case, the profile reconstruction technique is equally accurate for recognition of both small- and large-scale defects as the noise-produced patterns are of the same magnitude on all scales. One can rely on this technique, given the noise in the input signal (measurements) can be kept small enough. For the case of $f=H_j$, the noise-produced patterns decay for longer scales and, therefore, the reconstruction procedure is more accurate. Even for strong noise the recognition of sufficiently broad defects can be accurate. However, simultaneously with good resolution of large-scale defects, reliable identification and characterization of narrow defects become problematic or even impossible (it is impossible as well within the framework of the virgin equation system, without large-scale approximation). Comparing Eqs.\,(\ref{eq208}) and (\ref{eq215}), one can find the ``reference'' scale $L_{x\ast}$; measuring $f=(\partial H_z/\partial x)$ is preferable for defects small compared to this scale, while for a  larger scale of defects the reconstruction is more accurate with $f=H_z$;
\begin{equation}
L_{x\ast}=\frac{2\sqrt{5}}{\sqrt{3}}\sqrt{\frac{\langle(\varepsilon_{f=H_z}^\mathrm{(meas)})^2\rangle}
{\langle(\varepsilon_{f=\partial H_z/\partial x}^\mathrm{(meas)})^2\rangle}}\,.
\label{eq301}
\end{equation}

Accuracy of the reconstruction procedure is influenced not only by defect scale but also by its shape. One can rewrite Eqs.\,(\ref{eq208}) and (\ref{eq215})--(\ref{eq217}) as
\begin{equation}
a_\sigma^{(\partial{H_z}/\partial{x})}
=\gamma_{(\partial H_z/\partial x)}\frac{\sqrt{\langle(\varepsilon_f^\mathrm{(meas)})^2\rangle\,h_x\,h_y}}{\calH}\,,
\label{eq302}
\end{equation}
\begin{equation}
a_\sigma^{(H_j)}
=\gamma_{(H_j)}\frac{\sqrt{\langle(\varepsilon_f^\mathrm{(meas)})^2\rangle\,h_x\,h_y}}{\calH\sqrt{S}}\,,
\label{eq303}
\end{equation}
with form-factor $\gamma$
\begin{align}
\gamma_{(\partial{H_z}/\partial{x})}
&=\frac{\pi}{2\sqrt{5}}\frac{L_x}{L_y}\sqrt{\frac{L_x}{L_y}+\frac{L_y}{L_x}}\,,
\label{eq304}
\\[5pt]
\gamma_{(H_z)}&=\frac{\pi}{\sqrt{3}}\frac{L_x}{L_y}\sqrt{1+\frac{L_y^2}{L_x^2}}\,,
\label{eq305}
\\[5pt]
\gamma_{(H_x)}&=\frac{\pi}{2\sqrt{5}}\frac{L_x}{L_y}\left(\frac{L_x^2}{L_y^2}+1\right)\,,
\label{eq306}
\\[5pt]
\gamma_{(H_y)}&=\frac{\pi}{3}\left(\frac{L_x}{L_y}+\frac{L_y}{L_x}\right)\,.
\label{eq307}
\end{align}
In Table~\ref{tab} the behaviour of the form-factor for defects with similar dimensions in the $x$- and $y$-directions ($L_x\sim L_y$), defects stretched along the $y$-axis ($L_x\ll L_y$) and along the $x$-axis ($L_x\gg L_y$) is summarized. Notice, for $L_y\gg L_x$, the form-factor is diminished for $f=(\partial{H_z}/\partial{x})$ and $f=H_x$, and increased for $f=H_y$. For $L_x\gg L_y$, the form-factor is increased for all cases of $f$.

\setcounter{equation}{0}
\section{Conclusion}
\label{sec:conclusion}
We have developed an approach for the analytical calculation of noise-induced patterns produced when employing a complex non-local transformation for data reconstruction. This approach has been applied to the magnetic flux leakage (MFL) method for inspection of wellbore casing integrity; with the MFL method, the ferromagnetic layer thickness profile is reconstructed from measurements of the magnetic field above the layer. Within the framework of large-scale approximation, Eqs.\,(\ref{eq208}), (\ref{eq215})--(\ref{eq217}) have been derived for the device designs based on measuring $(\partial H_z/\partial x)$, $H_z$, $H_x$ and/or $H_y$, respectively; these equations form one of the principle results of our work. In particular, the case of $(\partial H_z/\partial x)$ has been found to be equally effective for all scales of defect, while use of $H_z$, $H_x$ and $H_y$ provide more reliable resolution of broader defects, but are inaccurate for the detection of narrow defects. The procedures of reconstruction of the ferromagnetic layer thickness from $H_z$, $H_x$ and $H_y$ are not equally sensitive to noise. More importantly, for these three cases, the sensitivity depends differently on the defect shape. This dependence is characterised by the form-factor, which, as one can see from Eqs.\,(\ref{eq305})--(\ref{eq307}), depends on the long-axis orientation for scar-shape defects. Thus, the combined usage of these procedures can be beneficial for the accuracy of recognition of defects.


\end{document}